\documentclass[pdflatex,sn-mathphys]{sn-jnl}
\usepackage{caption}
\usepackage[algo2e,ruled,linesnumbered]{algorithm2e}
\SetKwRepeat{Do}{Repeat}{Until}
\begin{document}

\title[AutoAssign+]{AutoAssign+: Automatic Shared Embedding Assignment in Streaming Recommendation}

\author[1]{\fnm{Ziru} \sur{Liu}}\email{ziruliu2-c@my.cityu.edu.hk}
\equalcont{These authors contributed equally to this work.}

\author[1]{\fnm{Kecheng} \sur{Chen}}\email{kechechen4-c@my.cityu.edu.hk}
\equalcont{These authors contributed equally to this work.}

\author[1]{\fnm{Fengyi} \sur{Song}}\email{fengysong2-c@my.cityu.edu.hk}

\author[2]{\fnm{Bo} \sur{Chen}}\email{chenbo116@huawei.com}

\author*[1]{\fnm{Xiangyu} \sur{Zhao}}\email{xianzhao@cityu.edu.hk}

\author[2]{\fnm{Huifeng} \sur{Guo}}\email{huifeng.guo@huawei.com}

\author*[2]{\fnm{Ruiming} \sur{Tang}}\email{tangruiming@huawei.com}

\affil[1]{\orgname{City University of Hong Kong, Kowloon Tong, Hong Kong}}
\affil[2]{\orgname{Huawei Noah's Ark Lab, Shenzhen, China}}




\abstract{In the domain of streaming recommender systems, conventional methods for addressing new user IDs or item IDs typically involve assigning initial ID embeddings randomly. However, this practice results in two practical challenges: (i) Items or users with limited interactive data may yield suboptimal prediction performance. (ii) Embedding new IDs or low-frequency IDs necessitates consistently expanding the embedding table, leading to unnecessary memory consumption. In light of these concerns, we introduce a reinforcement learning-driven framework, namely AutoAssign+, that facilitates Automatic Shared Embedding Assignment Plus. To be specific, AutoAssign+ utilizes an \textit{Identity Agent} as an actor network, which plays a dual role: (i) Representing low-frequency IDs field-wise with a small set of shared embeddings to enhance the embedding initialization, and (ii) Dynamically determining which ID features should be retained or eliminated in the embedding table. The policy of the agent is optimized with the guidance of a \textit{critic network}. To evaluate the effectiveness of our approach, we perform extensive experiments on three commonly used benchmark datasets. Our experiment results demonstrate that AutoAssign+ is capable of significantly enhancing recommendation performance by mitigating the cold-start problem. Furthermore, our framework yields a reduction in memory usage of approximately 20-30\%, verifying its practical effectiveness and efficiency for streaming recommender systems.}

\keywords{Recommender Systems, Reinforcement Learning, Cold-Start, Streaming Recommendation}



\maketitle

\section{Introduction}\label{sec1}
\begin{figure}[h]%
\centering
\includegraphics[width=0.65\textwidth]{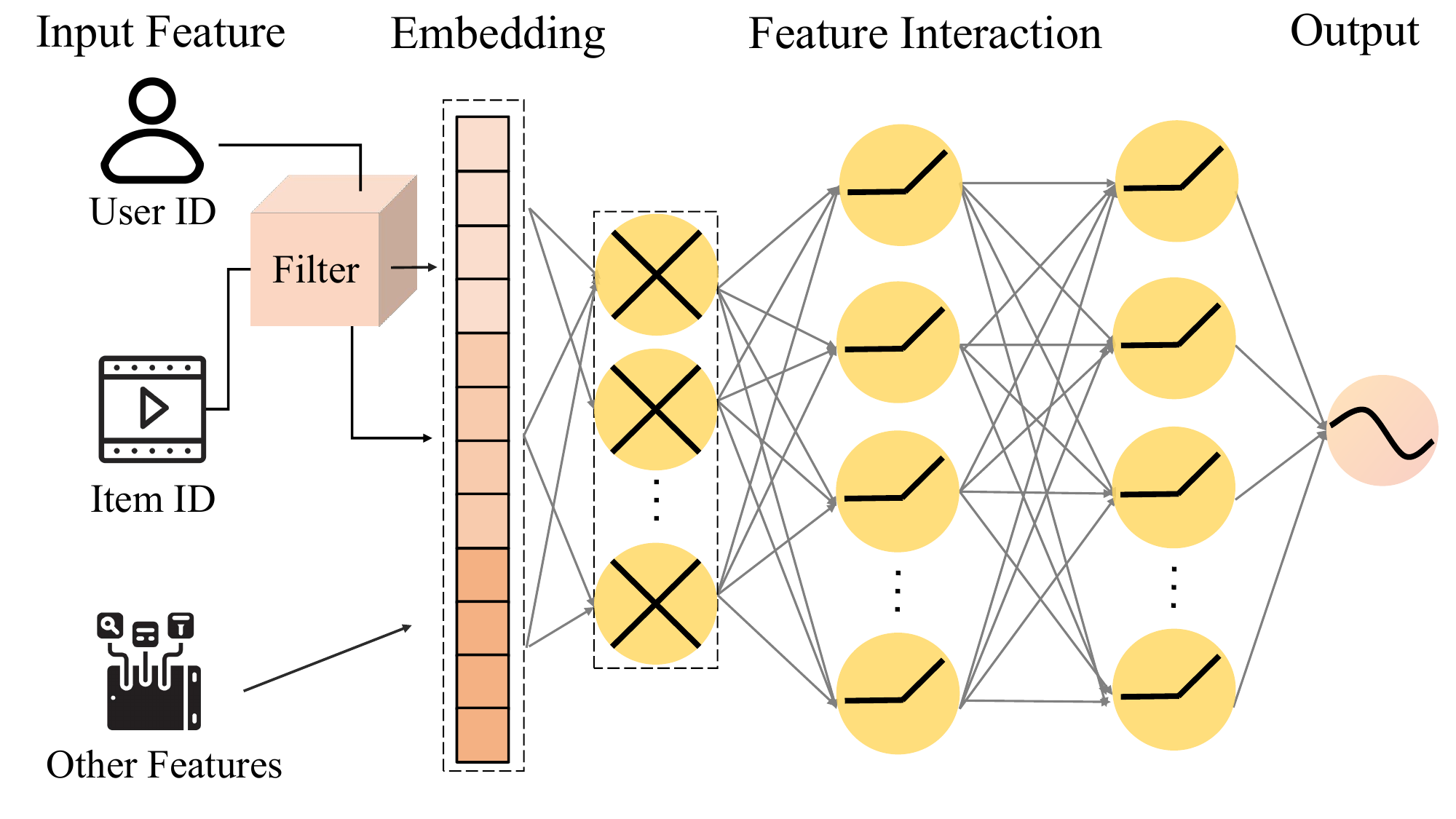}
\caption{A typical deep learning based recommendation model.}\label{fig1}
\vspace{-5mm}
\end{figure}

With the rapid growth of personalized online applications, recommender systems have been widely implemented by various online businesses, including E-commerce websites, news platforms, online advertising, and so on~\cite{an2019neural,ren2016user}. 
Among them, streaming recommendation~\cite{guo2019streaming,10.1145/3543507.3583237} is one of the common forms of recommender systems, where streaming data are constantly flowing into the recommendation models for training, thus better modeling the user's current preferences. In addition, streaming recommendations are particularly important for time-sensitive items, such as news, as they allow for rapid identification and distribution of relevant content to interested users, which is critical for commercial information retrieval systems.
Due to the ability to effectively capture the highly nonlinear relationship between user and item end-to-end, neural network-based models are rapidly becoming the mainstream of recommender systems.
As shown in Fig.~\ref{fig1}, existing deep recommendation models typically follow the ``Embedding \& Feature Interaction" paradigm~\cite{autodis}. The embedding layer serves as the encoder to represent sparse features in dense latent space, while the feature interaction layers serve to capture interactive signals among these features.

In a streaming recommender system, new items and users are continually added to the data corpus, creating a highly dynamic streaming environment that presents several challenges, which can be summarized as:
\begin{itemize}
\item \textbf{Cold-start}: The streaming recommender system is confronted with a constant influx of new users, many of whom can be classified as visitor-type users and possess extremely limited behavior information. Furthermore, the system is constantly updated with new items, yet there has not been enough interaction with these items to generate an adequate level of training data. The consequence of employing insufficiently trained new user/item embeddings is a significant decline in the performance of the recommendation model.
\item \textbf{Interval interaction}: Users' preferences and activity levels exhibit temporal fluctuations, manifesting as periods of ``\textit{active} $\rightarrow$ \textit{inactive} $\rightarrow$ \textit{active} $\rightarrow$ \textit{...}'' behavior. Similarly, items display a comparable trend over time. In order to predict the reactivation of users/items accurately, it is crucial to maintain the inactive user/item ID parameters updated with the recommendation model in real-time. Additionally, effectively recycling ID parameters of long-term inactive users/items can prevent the model from becoming overburdened with redundant data, leading to a significant reduction in model size and conserving space.
\end{itemize}

The challenges faced by streaming recommender systems can be attributed to the handling of (temporary) low-frequency user/item ID features. It is impractical and problematic to consider high and low-frequency IDs as equal entities. In industrial streaming recommender systems, a typical approach to address low-frequency IDs is to replace their embeddings temporarily with shared embeddings.
As shown in Fig.~\ref{fig1}, a pre-defined frequency threshold can be used to filter those low-frequency IDs before the embedding lookup. However, this straightforward approach has a few obstacles when applied to streaming recommendation systems.: (i) Determining optimal pre-defined thresholds for user/item IDs requires human expert knowledge and can be time-consuming; (ii) The distribution of IDs can change over time, resulting in an inadequate performance when a fixed threshold is used. One potential solution is to use an adaptive threshold that can be adjusted dynamically, based on the current distribution of IDs.

Recent research efforts have focused on addressing the cold-start problem in streaming recommender systems using auxiliary information~\cite{cold1,cold2,cold3}. Additionally, meta-learning has emerged as a popular approach for learning global and general information for relevant tasks, thereby facilitating the initialization of new, related tasks. For example, MetaEmb~\cite{meta1} generates an initial ID embedding for new items using item features, while MWUF~\cite{meta4} employs the average of all existing ID embeddings in conjunction with scaling and shifting functions to initialize cold-start IDs.
However, the auxiliary information methods~\cite{cold1,cold2,cold3} rely on rich user profiles or item attributes, such as user social networks or item images, which may not be available in scenarios involving only user/item IDs. Moreover, meta-embedding-based approaches that use global parameters in the feature interaction layer or pre-trained parameters of the embedding table to enhance initialization for new IDs can introduce deviations in the embeddings of high-frequency IDs.

To tackle the aforementioned difficulties, 
we propose Automatic Shared Embedding Assignment Plus (\textbf{AutoAssign+}), a reinforcement learning-based approach, which utilizes an \textbf{Identity Agent} to dynamically and field-wisely assign shared embedding and unique embedding to the users/items ID features that appear anytime in the streaming recommendation scenario. Besides, one \textbf{critic network} is applied to optimize and fine-tune the
policy of the agent based on TD error. In our previous version, the Identity Agent of \textbf{AutoAssign} generates the candidate embeddings based solely on the input features and doesn't take into account the performance of the recommendation model. In AutoAssign+, the critic network is introduced to optimize the Identity Agent by evaluating the quality of the candidate embeddings generated by the Identity Agent, allowing it to adjust and optimize its candidate embedding generation process to improve the recommendation model's performance. Notably, our Automatic Shared Embedding framework is designed to be "plug and play," allowing it to be applied to various deep recommendation models.

The main contributions of our work can be concluded as:

\begin{itemize}
    \item We propose a new framework AutoAssign+ for streaming recommendation scenarios, which is compatible with various deep recommendation models to alleviate the low-frequency ID problem;
    
    \item AutoAssign+ utilizes an actor-critic structure from reinforcement learning along with hierarchical shared embeddings to dynamically assign optimal embeddings for each user/item ID based on their occurrence frequency.
 
    \item We performed extensive experiments on popular datasets to demonstrate the effectiveness and superiority of AutoAssign+. The results show that AutoAssign+ significantly outperforms the performance of AutoAssign while reducing storage space by 20\%-30\%.
\end{itemize}

\begin{figure*}[h]
  \centering
  \vspace{-6mm}
  \includegraphics[width=0.9\linewidth]{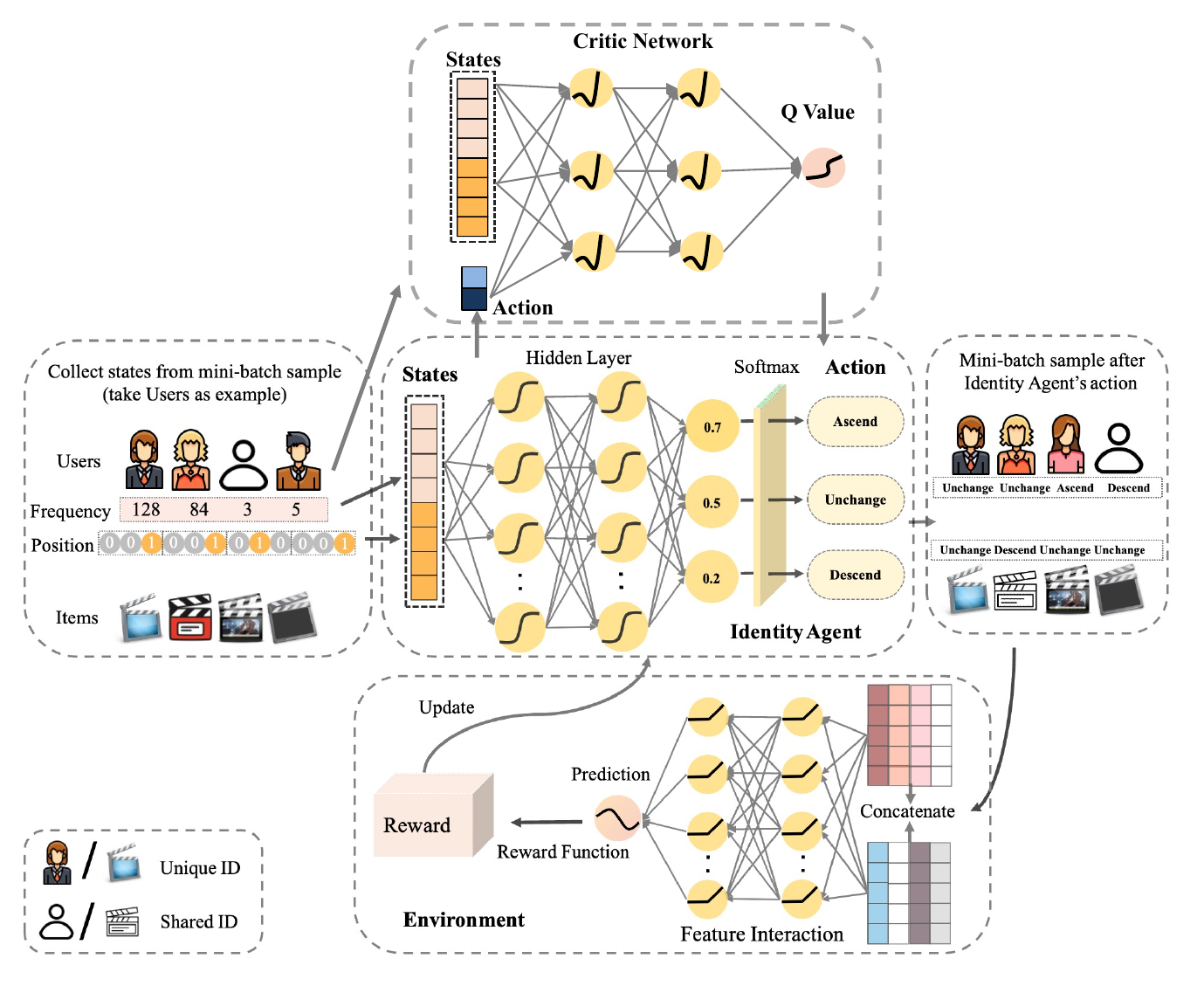}
  \caption{The proposed overall framework contains an Identity Agent, which serves to assign a shared ID to the User and Item before the embedding layer.}
  \label{fig:framework}
  \vspace{-5mm}
\end{figure*}

\section{Preliminary}
\subsection{Deep Recommendation Model}
The deep recommendation model utilizes dense vectors (embeddings) of instances as input and makes predictions by feeding those embeddings into subsequent feature interaction components \cite{DBLP:journals/kais/LiuTZW23,DBLP:journals/kais/HeWDS23}. The instance dense vectors can be constructed by concatenating the vector representations of each sparse (categorical) feature \cite{ctr}. As illustrated in Fig.~\ref{fig1}, for a given user-item pair, their IDs and corresponding features can be represented as:
\begin{equation}
\begin{split}
   \boldsymbol{x} = [\boldsymbol{x}^{u}, \boldsymbol{x}^{i}, \boldsymbol{x}_{1},\boldsymbol{x}_{2},...,\boldsymbol{x}_{n}], 
\end{split}
\label{1}
\end{equation}
where $\boldsymbol{x}_{u}$ and $\boldsymbol{x}^{i}$ are feature vectors for user and item, $n$ is the number of other feature fields and $\boldsymbol{x}_{i}$ is the high-dimensional sparse vector of $i^{th}$ field. Then embedding layer transforms those sparse vectors into dense vectors:
\begin{equation}
\begin{split}
   \boldsymbol{e}_{i} = \boldsymbol{W}_{i}\boldsymbol{x}_{i},
\end{split}
\label{2}
\end{equation} 
 where $\boldsymbol{W}_i \in \mathbb{R}^{d \times h}$ is a matrix (embedding table), $d$ is the embedding dimension of dense vector and $h$ is the number of feature value in $i^{th}$ field. These embeddings are feed-in into the subsequent feature interaction layer to extract informative interactions. In the binary classification task, the predicted target is $y \in{\{0,1\}}$ where $0$ refers to dislike and $1$ refers to like in explicit user behavior. We assume that the parameters of the deep recommendation model are $\theta$. Depending on the feature interaction function $f(\cdot)$, we get the prediction $\hat{y}$:
\begin{equation}
\begin{split}
   \hat{y} = f(\theta,\boldsymbol{e}^u,\boldsymbol{e}^i,\boldsymbol{e}_n),
\end{split}
\label{3}
\end{equation} 
where $\boldsymbol{e}^u$ and $\boldsymbol{e}^i$ respectively denote the user ID and item ID embedding, and $\boldsymbol{e}_n$ is the concatenate of $n$ number other feature fields. The typical optimization target for binary classification tasks is Logloss or Mean Square Error.

In the context of deep recommendation models, user/item ID embeddings are crucial as they encode the latent features of specific users/items \cite{DBLP:journals/kais/DuTC23,DBLP:journals/kais/ChenZC23}. However, in streaming recommendation scenarios, insufficient training of ID embeddings can arise due to two situations. Firstly, the emergence of new users and items is commonly referred to as the cold-start problem. Secondly, certain items or visitor-like users may have an inherently unpopular nature, resulting in a long-tailed distribution issue. The initialization of low-frequency and unpopular ID embeddings with random values cannot provide valid information, ultimately leading to poor recommendation performance.

\subsection{Low-Frequency Filter}
An intuitive approach to tackle the previously mentioned issue is to temporarily use a shared embedding to represent the low-frequency IDs. As shown in Fig.~\ref{fig1}, a frequency threshold is adopted in a real-world industry recommender system to filter those IDs with a frequency lower than the threshold before performing the embedding lookup. Specifically, suppose the frequency threshold for User ID is $\tau$ and we set a shared ID~$ID_{shared}$, the User ID is collected as $(ID_i, F(i))$, where $i$ represents the $i^{th}$ user and $F(\cdot)$ represent its current occurrence frequency. In low-frequency filter:

$$(ID_i,F(i)) = \left\{
\begin{aligned}
ID_i & ~~if ~F(i) > \tau \\
ID_{shared} & ~~if ~F(i) \le \tau 
\end{aligned}
\right.,$$

\noindent By training a shared embedding $ID_{shared}$ on a large number of low-frequency IDs, it can obtain abundant generalized information. However, this method has some practical drawbacks. Firstly, determining a fixed threshold $\tau$ to identify low-frequency IDs requires expert knowledge or extensive search. Secondly, the frequency distribution of different ID fields (such as user ID and item ID) varies spatially and temporally, making it sub-optimal to use a fixed threshold to define low-frequency IDs.

\section{Framework}
The AutoAssign+ framework is designed to address the limitations of existing methods by automatically assigning user/item embeddings to low-frequency targets, which helps to alleviate the negative impact caused by the low-frequency ID features on the recommendation performance. In this section, we will provide a detailed description of our proposed approach.

\subsection{Reinforment Learning Setting}
To address the embedding assignment problem, we formulate it as a Markov Decision Process (MDP) that can be analyzed within the reinforcement learning framework where we train the Identity Agent to make optimal embedding assignment decisions.
\subsubsection{State}
The entirety of the deep recommendation model's pipeline is considered to be the environment. With the available data sample obtained from the stream, the deep recommendation model generated an output using the Identity Agent's policy decision. Under this pipeline, we define a group of hierarchical candidate shared ID with size $k$, i.e., $\{S_1, S_2,..., S_k\}$, and its corresponding shared embedding table $\boldsymbol{E}_{shared}= [{e_{s1},e_{s2},...,e_{sk}}] \in \mathbb{R}^{d \times k}$, where $d$ is the embedding dimension of shared ID and $k$ is the size of shared ID group. In the case of this study, we choose $k=2$. These shared embeddings are pre-trained from the same data and parameters randomly. This hierarchical shared embedding setting acts as a buffer before the unique embedding is assigned to a specific user/item in the recommendation process. 


The state is characterized as the present frequency of ID features along with their current position $p_i$, within the shared ID group of candidates (i.e., the $i^{th}$ embedding vector of the shared embedding table), $State = (F(\cdot), p_i)$, where $i \in [1,k+1]$ and $F(\cdot)$ is the current occurrence frequency of a certain ID. Note that: (i) there is an additional position $k+1$, which refers to a unique ID already assigned to a certain ID, (ii) the state of a new coming ID is initialized as $State_{init} = (1, 1)$, which means that its current frequency is one and assigned with the first embedding vector in the shared embedding table. 

\begin{figure}[h]
\vspace{-2mm}
  \centering
  \includegraphics[width=0.95\linewidth]{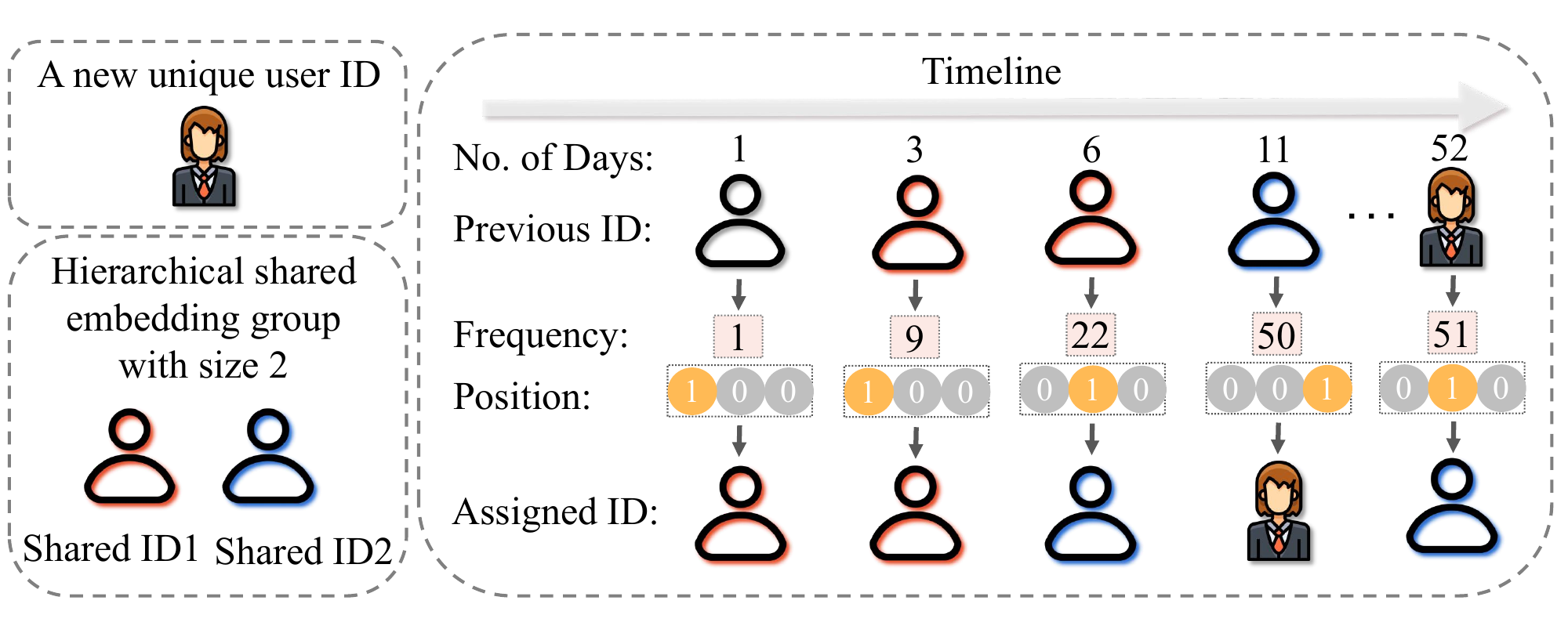}
  \caption{Assign a shared ID / unique ID to a new user by AutoAssign+.}
  \label{fig:process}
  \vspace{-7mm}
\end{figure}

\subsubsection{Action}
The design and architecture of the Identity Agents are identical for both the user and the item. We use the user ID agent to illustrate the action as shown in Fig. \ref{fig:framework}. Given the state of a user ID, $(F(\cdot), p)$, we encode the frequency into a dense representation $\boldsymbol{e}^{f}$ and convert the discrete position of ID into one-hot vectors $\boldsymbol{v}$ with size $k+1$. After that, these two transformed representations are concatenated and fed into multilayer perceptrons (MLP) with $m$ 
hidden layers and activation function $tanh$:
\begin{equation}
\begin{aligned}
   h_1 & = tanh(\boldsymbol{W_1}[\boldsymbol{e}^{f} : \boldsymbol{v}]) + b_1 \\
   h_2 & = tanh(\boldsymbol{W_2}h_1) + b_2  \\
   & ... \\
   h_m & = tanh(\boldsymbol{W_{m}}h_{m-1}) + b_{m} \\
   logits & = softmax(h_m)
\end{aligned}
\label{4}
\end{equation} 
Subsequently, the hidden state of the last layer denoted as $h_m$ is further processed by 
the Softmax layer and outputs three probabilities that correspond to the actions of "Ascend", "Unchanged", and "Descend". Specifically, selecting "Ascend" as an action would indicate raising the position of the ID within the shared embedding candidates set (e.g., from $p_i$ to $p_{i+1}$). ``Unchanged" means to remain in the same position. ``Descend" means the Identity Agent decides to make a ``rollback" on the position of the given ID (e.g., from $p_{i+1}$ to $p_{i}$).
Recall that the group of candidate shared IDs is hierarchical, then the motivation for designing the action of ascending or descending movements in this hierarchical group is shown in Fig.  \ref{fig:process} The approach adopted for updating the position of an ID within the shared embedding candidates set is as follows: (i) when the cumulative frequency of a particular user ID reaches a specific threshold, the ID is elevated to a higher-frequency representation level; (ii) if an ID's popularity declines, it will descend to a lower-frequency representation. It is noteworthy that (i) an ID's position shifting from $p_k$ to $p_{k+1}$ indicates that it will no longer be a shared ID, and instead, it will be assigned its own ID and corresponding embedding, which will be initialized using the last shared embedding $e_{sk}$. This is based on the intuition that the embedding of shared IDs is generally well-trained and, therefore, possesses sufficient generality to provide suitable initialization for individual IDs. (ii) When certain IDs move from unique IDs to shared IDs, it helps in saving storage and avoiding parameter updating issues.

\subsubsection{Reward}
The primary objective of the Identity Agent is to enhance the recommendation performance by dynamically assigning either unique IDs or shared IDs to ID features that are present in the data stream. To this end, we propose employing a reward function that can effectively evaluate the model's performance. Specifically, we define the reward as the prediction loss of the recommendation model utilized in our experiment. Given a pair of user $u$ and item $i$ and their current prediction loss $L$ and their corresponding last $T$ prediction losses $L^u = (L^u_1,L^u_2,..., L^u_T)$ , $L^i = (L^i_1,L^i_2,..., L^i_T)$, where $L^{u/i}_t$ refers to the $t^{th}$ prediction loss of user $u$ / item $i$, the reward is then defined as the difference between the current loss $L$ and the average of last $T$ prediction losses: 

\vspace{-3mm}
\begin{align}
\label{5}
   R^{ u} & = \frac{1}{T}\sum_{t=1}^{T}{L^{u}_{t}} -  L \\
   R^{ i} & = \frac{1}{T}\sum_{t=1}^{T}{L^{i}_{t}} -  L 
\end{align}

The policy of the Identity Agent can be thought of as maximizing the reduction in the present prediction loss concerning the previous average loss. In this paper, we set $T=30$, which corresponds to the prediction loss of the last 30 days. Using the past $T$ prediction losses instead of just the last one has several benefits in the proposed approach. It allows the system to consider the historical performance of the model and incentivizes it to maintain a consistent level of accuracy over time. Also, using the past T prediction losses can help to reduce the impact of noise or fluctuations in the loss function that may occur due to random variation in the input data or other factors. This approach allows the Identity Agent to make more informed decisions that result in lower prediction loss during the process of continuous learning and optimization.

\subsection{AutoAssign+}
The overall framework of AutoAssign+ includes two steps in the learning process, as shown in Fig. \ref{fig:framework}:
\begin{itemize}
    \item \textbf{The Forward Step.} 
    In the first step, the Identity Agent receives the corresponding states of each incoming ID and takes action to assign either a unique ID or one of the hierarchical shared IDs to each ID in a batch of user-item interactions. The action value and user-item combined features are further processed by the critic network for calculating the Q-value, which can fine-tune the performance of the Identity Agent. After the ID embeddings are assigned, the user and item IDs are processed using the embedding layer and concatenated, following which they are fed into the inference layer. The predicted output is then compared to the actual output, and the mean square error (MSE) loss function is used to calculate the overall loss.
    
    \item \textbf{The Backward Step.} We first update the critic network based on the gradients of Q-value and TD error until its convergent. Then the recommendation model parameters are updated based on the computed loss, and the Identity Agent parameters are updated by determining the reward based on the prediction loss. As a result, during continuous training and evaluation, the data-driven Identity Agent is fine-tuned and becomes capable of making more informed decisions.
\end{itemize}


\subsubsection{Critic Network}

\begin{figure}[ht]
    \vspace{-4mm}
    \flushleft
    \centering
    \includegraphics[width=0.66\linewidth]{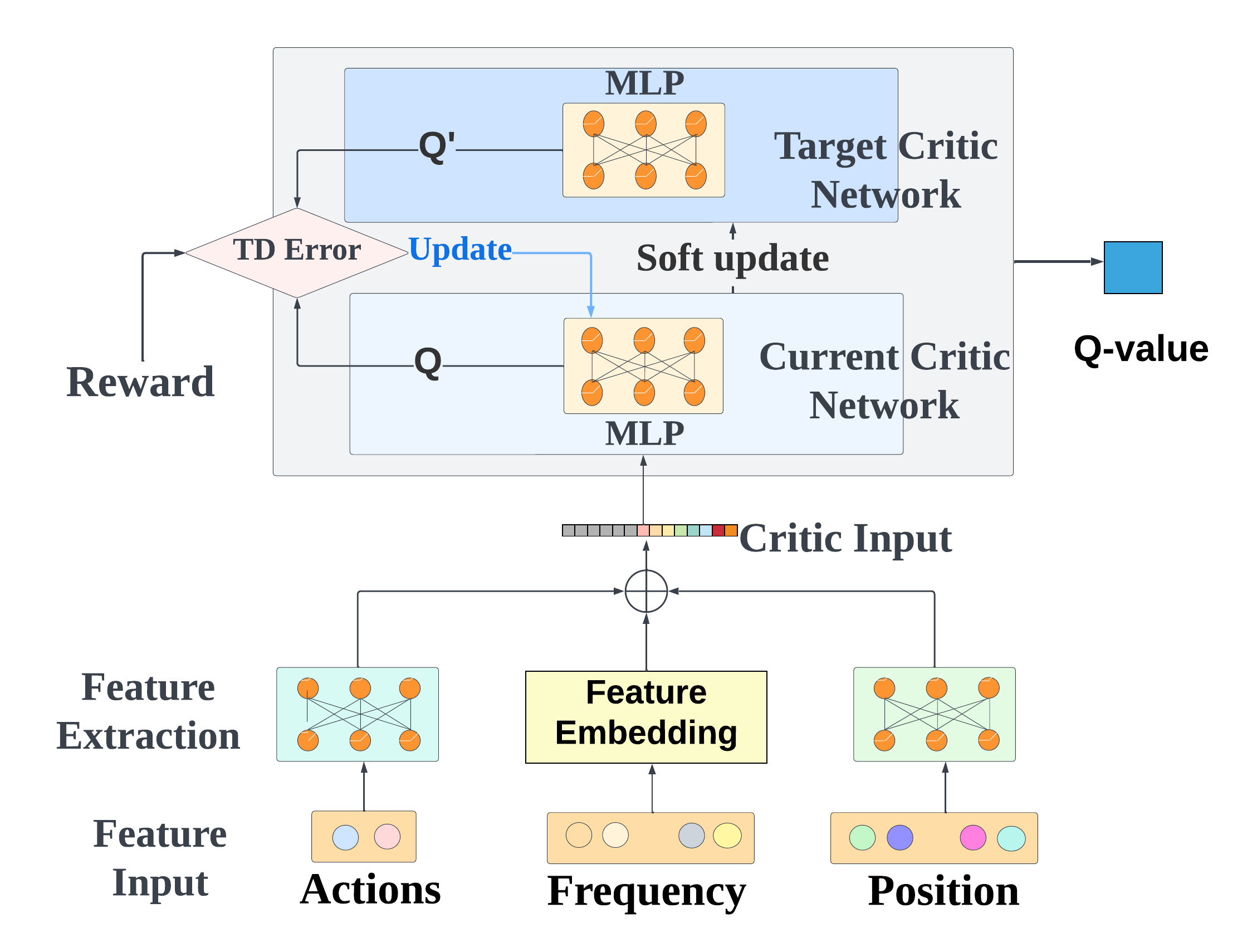}
    \caption{Structure of Critic Network.}
    \label{fig:Critic}
    \vspace{-6mm}
\end{figure}

The standard approach involves using a critic network to estimate the action value generated by the actor network, and then updating the actor network parameters based on this value. However, when using the Identity Agent as the actor network, the challenge lies in designing a suitable structure for the critic network to facilitate parameter updating. Our solution involves a shared-bottom layer in the critic network that simultaneously transforms the user-item features and action information. To achieve this, we first apply an embedding layer and an MLP structure to extract the features and then combine the resulting user-item feature and action information as input to a differentiable action value network that is parameterized by $\phi$. This network outputs the estimated Q-value based on the state-action information. Specifically, given the current state $s$ and action $a$, the Q-value is calculated as follows:
\begin{align} 
\label{Q}
   Q(s,a)
   &= \mathop{\mathbb{E}}[R+\gamma V(s') \mid s,a] \\
    &= R+\gamma \sum_{s' \in \mathcal{S}} p_{s,a, s'} \cdot \max Q(s',a')
\end{align}

where $V(.)$ is the state value function, and $s',a'$ is the state and action value of next step. The hyperparameter $\gamma$ is the discount rate set as 0.95 for our case. In our case, the action value $a$ is estimated by the Identity Agent, and the next state $s'$ is determined with a probability equal to 1. Therefore, the Q-value function in a multi-critic structure can be calculated as follows:
\begin{eqnarray} \label{Q}
Q(s,a;\phi)
&=& R +\gamma Q(s',a';\phi) 
\end{eqnarray}
Since the actor-critic network framework often faces the challenge of failing to converge, to address this issue, we incorporate the concept of Deterministic Policy Gradient Algorithms \cite{silver2014deterministic} by integrating target networks into the learning framework. The target networks share the exact same structure as the critic networks that have been proposed. We denote it as $Q(s,a; \mathbf{\widetilde{\phi}})$, which have lagging parameter $\widetilde{\phi}$. The structure of critic network is shown in Fig. \ref{fig:Critic}.

\subsection{Overall Optimization}
In this section, we will discuss the optimization process of the deep recommendation model, Identity Agent, and critic network, and present the overall optimization framework under the streaming recommendation setting.

\subsubsection{Critic Network Updates.}
$\phi$ is the crucial parameter in the critic network, which determines the action Q-value. Given transition. Given the action and state information, the TD target of the target critic network is derived from:
    \begin{equation}
        TD = R + \gamma Q(s',a'; \widetilde{\phi})
    \end{equation}
    where $a' = \pi(s';\mathbf{\widetilde{\theta}}_k)$ is the estimated next action from target Identity Agent.
    The Q-value generated from the current critic network, which estimates the current action value is defined as:
    \begin{equation}
        Q = Q(s,a; \mathbf{\phi})
    \end{equation}
    
\noindent We calculate the average TD error $\delta$ after the training among the batch:
\begin{small}
    \begin{eqnarray} \label{TD}
        \delta 
        &=& \frac{1}{b} \sum\limits_{k=1}^b (TD_k - Q_k) \\
        &=& \frac{1}{b} \sum\limits_{k=1}^b [R_k + \gamma Q_k(s',a'; \widetilde{\mathbf{\phi}})-Q_k(s,a; \mathbf{\phi})] \nonumber 
    \end{eqnarray} 
\end{small}

Then we update the current critic network for each task by the following gradient decent method with a
learning rate $\alpha_{\mathbf{\phi}}$
    \begin{equation} \label{cu}
        \phi \gets \phi - \alpha_{\mathbf{\phi}}\mathbf{I}\delta \nabla_{\mathbf{\phi}}Q(s,a; \mathbf{\phi})
    \end{equation}
    where $\nabla_{\mathbf{\phi}}Q(s,a; \mathbf{\phi})$ is the gradient of the target Q-value. This completes the optimization of the current critic networks.

The target critic network is updated until the current critic network reaches the convergence condition towards the direction of parameters in current networks:
    \begin{equation}
    \begin{split}
        \widetilde{\phi} = \beta \widetilde{\phi} + (1-\beta)\phi
    \end{split}
    \end{equation}
    where $\beta \in [0,1]$ is the soft update rate.

\subsubsection{Identity Agent Updates.}
Let $\theta$ represent the parameters in the feature interaction model and $\omega$ represent the parameters in the embedding table. In a binary classification task, the Mean Square Error (MSE) can serve as the optimization target. Suppose we have a mini-batch of user-item pairs $\{u_j,i_j\}^{N}_{j=1}$ and their corresponding labels $\{y_j\}^{N}_{j=1}$, where $y_j \in \{0,1\}$ and 0/1 denotes a negative/positive view about a certain item of a user. The MSE loss is defined as follows:
\begin{equation}
 MSE(\theta,\omega) = \frac{1}{N}\sum_{j=1}^{N}{(y_j-\hat{y}_j)}^2,
\end{equation} 
where the predicted label of the deep recommendation model is denoted as $\hat{y}_i$. 

With the Identity Agent parameterized by $\Omega$ and the reward function designed in Eq. (\ref{5}), the Identity Agent maximizes the expectation of rewards as follows:

\begin{equation}
F(\Omega)=\mathbb{E}_{a \Omega(a \mid s)}(a \mid s)
\label{6}
\end{equation}

where $a$ is the action and $s$ is the state. Before the TD error $\delta$ converges to threshold $\epsilon$, we update the Identity Agent parameterized by $\Omega$ through the gradients back-propagation of loss function for each layer after the forward process of each batch transitions:
    \begin{equation} \label{iu}
        \Omega \gets \Omega + \alpha_{\Omega} \mathbf{I} \nabla_{\Omega} \mathcal{J}(\Omega)
    \end{equation}
    where the loss for tower layers is defined by the negative of average Q-value $\mathcal{J}(\Omega) = - \frac{1}{b}\sum\limits_{k=1}^b Q'(s, \Omega(a\mid s))$, which is generated from the critic network.

After the convergence of the critic network, we tend to optimize the Identity Agent based on the overall objective. Practically, it isn't easy to calculate the exact value of expectation reward, thus we use the Monte-Carlo sampling \cite{mcs} to estimate its gradient along with the optimization algorithm of Eq. (\ref{6}):

\begin{align}
   \nabla_{\Omega} F(\Omega) &= \sum_{a}{R(a\mid s)\nabla\Omega(a\mid s)}\\
   &= \sum_{a}{R(a \mid s)\Omega(a\mid s)\nabla log \Omega(a\mid s)}\\
   &= \mathbb{E}_{a~\Omega(a\mid s)}[R(a\mid s)\nabla log \Omega(a\mid s)] \\
   & \approx \frac{1}{N} \sum_{i=1}^{N}{R(a\mid s)\nabla log \Omega(a\mid s)}
\end{align}

where $N$ is the number of samples. After obtaining the gradient of rewards, the parameter of the Identity Agent network can then be updated with a learning rate $\alpha_{I}$ by:
\begin{equation}
   \Omega \gets \Omega + \alpha_{I} \nabla_{\Omega}F(\Omega).
\label{9}
\end{equation}

\begin{algorithm}[ht]
  \SetAlgoLined
	\KwData{Boundless data stream $S$ in recommender system in the form of (userID, itemID, ground- truth label) with a mini-batch size $b$: $D = \{(u_j,i_j,y_j)\}_{j=1}^{b}$; Recommendation model parameter $\Theta$; Identity Agent parameter $\Omega$; Critic network parameter $\phi$}
	\KwResult{Fine-tuned parameter of recommendation model, Identity Agent, and critic network $\Theta^{*}$, $\Omega^{*}$ and $\phi^{*}$}

\BlankLine
	Initialize $\Theta$, $\Omega$ and $\phi$ from kaiming initialization \cite{kaiming};
	
	\Do{
	$\Theta$ \& $\Omega$ converge or encounter the end of the data stream
	}
		{
		Sample a validation batch from the last transaction in the history data stream: $D_{val} = {\{(u_j,i_j,y_j)\}^{b}_{j=1}}$ 
		
		Sample action $a_{val}~\Omega(D_{val})$ by importance sampling;

            Forward action $a_{val}$ and state information into the critic network, calculate the TD error based on Eq. (\ref{TD}). 
            Then updated parameter of the critic network by Eq. (\ref{cu}).

            \eIf{$\delta \geq \epsilon$}{
              Update the Identity Agent based on Eq. (\ref{iu})\;
              }{
		Adjust position of $(u_j,i_j)_{j=1}^{b}$  based on $a_{val}$ to get $(\hat{u}_j,\hat{i}_j)_{j=1}^{b}$ 
		
		Calculate the reward by Eq. (\ref{5}) using the adjusted $\hat{D}_{val}$
		
		Updated parameter of Identity Agent $\Omega$ by Eq. (\ref{9})
		
		Given the current batch of user-item data $D_{train} = {\{(u_j,i_j,y_j)\}^{b}_{j=1}}$ 
		
		Calculate the action: $a_{train} = ~argmax(\Omega(D_{train}))$
		
		Adjust position of $(u_j,i_j)_{j=1}^{b}$ according to $a_{train}$
		
		Updated the parameter of recommendation model $\Theta$ using the adjusted $\hat{D}_{train} = {\{(u_j,i_j,y_j)\}^{b}_{j=1}}$.

	}
        Perform soft updates of critic network
        $\widetilde{\phi} \leftarrow \beta \widetilde{\phi} + (1-\beta)\phi $}
\caption{Algorithm for overall optimization}
\label{algorithm1}
\end{algorithm}

Next, we present the optimization pipeline of the whole framework. The user-item intersection data are in-flowed boundlessly in the online recommender system in a data stream. We optimize the framework by iteratively collecting mini-batch data with a size of $b$ and updating the parameters of the recommendation model, Identity Agent, and critic network alternately.

As shown in Algorithm \ref{algorithm1}, in the initialization stage, we initialize $\Theta$, $\Omega$ and $\phi$ from the kaiming initialization \cite{kaiming} (line 1). The state of each ID feature that first time emerges in the data stream is initialized from (1,1). In the $\Omega$ updating stage, we sample the last batch in the historical transaction data as the validation batch (line 3) and sample action $a_{val}~\Omega(D_{val})$ using the importance sampling \cite{importance} (line 4).
The state and sample action information is further processed by the critic network to calculate Q-value and corresponding TD error $\sigma$, which updates the parameter of the current critic network $\phi$ (line 5). If the critic network is not converged to $\epsilon$, we update the Identity Agent by the critic network (lines 6-8). Next we calculate the adjusted the position $(\hat{u}_j,\hat{i}_j)_{j=1}^{b}$ based on sampled action $a_{val}$ (line 9). We then calculate the reward by Eq. (\ref{5}) with the ground-truth label under the evaluation mode of the recommendation model (line 10). Then the parameter of Identity Agent $\Omega$ can be updated by Eq. (\ref{9}) (line 11). Here we introduce the $\Theta$ updating stage: First, we collect the current batch of transaction data (line 12) and get the corresponding action (the highest probability among three actions) from the Identity Agent $a_{train}~argmax(\Omega(D_{train}))$ (line 13). Next we adjust and record the position of ${(u_j,i_j)}^{b}_{j=1}$ according to $a_{train}$ (line 14). Then the parameter of recommendation model $\Theta$ can be updated using the adjusted data $\hat{D}_{train} =\{(\hat{u}_j,\hat{i}_j,y_j)\}_{j=1}^{b}$ (line 15). The whole optimization process can be terminated when meeting the satisfied criteria or at the end of the data stream.

\section{Experiment}
In this section, we will provide a detailed description of our experimental setup. Subsequently, we will conduct extensive experiments on three datasets to investigate four research questions:
\begin{itemize}
\item[(1)]What is the overall performance of our method among different datasets, and whether we have solved the significant cold-start problem? 
\item[(2)]How is the generalizability of our method? Is it in line with the actual streaming recommendation in the industry?
\item[(3)]How does each component in our proposed method contributes to the final achievement? 
\item[(4)] Is our method able to allocate unnecessary memory usage? If the answer is affirmative, then what is the scale of specificity?
\end{itemize}

\subsection{Experiment Settings}
\subsubsection{Dataset} \label{data}
We conduct the overall performance experiments on three popular datasets in recommendations.\\ 
\textbf{MovieLens 25M}\footnote{\url{https://grouplens.org/datasets/MovieLens/}}: MovieLens 25M is a newly released stable benchmark dataset for personalized movie recommendations. It contains 25 million intersections and one million tag applications applied to 62,000 movies by 162,000 users. The ratings range from 1 to 5.\\
\textbf{MovieLens Latest}\footnote{\url{https://grouplens.org/datasets/MovieLens/}}:  Similar to MovieLens 25M, it contains 27,000,000 ratings and 1,100,000 tag applications applied to 58,000 movies by 280,000 users. The major difference between this Dataset and MovieLens 25M is that this dataset is highly sparse as it only has a density of 0.16\%, which is suitable for exploring the cold-start problem. \\
\textbf{Netflix Price}\footnote{\url{https://www.kaggle.com/datasets/netflix-inc/netflix-prize-data}}: Netflix price open competition dataset for predicting user ratings on movies. The movie rating file contains more than 100 million ratings from 480,000 anonymous Netflix users and more than 17,000 movies. Data are collected from October 1998 to December 2005.

In order to simulate the real-world streaming recommendation scenario, we sort the samples uniformly based on their timestamps and use one epoch from the start to the end without multiple iterations. It is important to construct the dataset through timestamps to align with the real-world recommendation scenario, where user interests and preferences are sequential and change over time, avoiding the issue of data traversal. After sorting, we use the first 80\% of the samples for parameter training and evaluate the accuracy and loss of the prediction results over the last 20\% of the data. In the testing phase, we predict and record the performance alternately and continuously train our model \cite{esapn}, as mentioned in the previous optimization section.

\begin{table}[h]
\renewcommand\arraystretch{1.3}
\begin{center}
\begin{minipage}{\textwidth}
\vspace{-3mm}
\caption{Overall information of Datasets}\label{data}
\centering
\begin{tabular}{ c c c c c c }
    \hline
        Dataset & Total User & Total Item & Ratings & Density \\ \hline
        MovieLens Latest & 280,000 & 58,000 & 27,000,000 & 0.16\% \\ 
        MovieLens 25M & 162,000 & 62,000 & 25,000,000 & 0.25\% \\ 
        Netflix Price & 480,000 & 17,000 & 100,000,000 & 1.22\% \\ \hline
\end{tabular}
\vspace{-5mm}
\end{minipage}
\end{center}
\end{table}

\subsubsection{Evaluation Metrics}
Since the focus of our study is on the binary classification task, we have converted the ratings of the three datasets to binary labels where 1 (${>}3$) represents the user likes the item and 0 (${\le}3$) represents the user dislikes the item. To evaluate the performance of our model, we have used Mean Square Error (MSE) loss, Accuracy, and Area Under the Curve (AUC) as the evaluation metrics.

\subsubsection{Implementation details}
In our experiments, we adopt one hidden layer in our agent network with a size of 512, and the embedding dimension of frequency is 32. For the recommendation model, we adopt an embedding size of 128 for both the user and item and two hidden layers with a size of 512 and use $LeakyRelu$ as the activation function. Both the agent networks and recommendation networks use Adam optimizer and 0.0001 and 0.001 as the learning rates, respectively. The structure of the critic network is an input embedding layer with dimension 128, a $128 \times 512 \times 256$ Multi-Layer Perceptron (MLP) as the bottom layer, and a $256 \times 128 \times 64 \times 1$ MLP as the tower layer applying Adam optimizer with learning rate 0.0001, the default soft update rate $\beta$ = 0.2 and batch size is 500.
According to the experimental results, we set the size of the user/item shared embedding candidates group to 1 and 2, respectively. To clearly demonstrate the effectiveness of our method and solely investigate the difference between well-trained and under-trained embeddings, we only utilize the user and item embeddings as the input of our recommendation model in the horizontal comparison of these datasets. The implementation code is available online to ease reproducibility.\footnote{https://github.com/Applied-Machine-Learning-Lab/AutoAssign-Plus}

\subsubsection{Baselines}

\begin{itemize}
\item \textbf{Origin} This model does not employ a threshold to filter out low-frequency ID features. It is used as a reference to demonstrate the negative impact and harm that low-frequency IDs and the cold-start problem can have on model performance.
\item \textbf{LFF (Grid/Random)} LFF method is used to filter low-frequency IDs by setting a fixed frequency threshold. Those IDs that occur less than this threshold are assigned one shared embedding. The best two thresholds for User and Item are selected through grid search and random search.
\item \textbf{AutoAssign} Automatic Shared Embedding Assignment method with only one agent network, which is the original version.
\end{itemize}

\subsection{Overall Performance}
Table \ref{overall} presents the overall performance of the evaluated methods. From the results, we can observe that

\begin{table}[h]
\renewcommand\arraystretch{1.3}
\begin{center}
\begin{minipage}{\textwidth}
\caption{Overall Performance}\label{overall}
\centering
\resizebox{\textwidth}{!}{%
\begin{tabular}{c c c c c c c}
    \hline
    \hline
        Dataset & Model & MSE Loss & Accuracy & AUC & Time (min) & Improve (\%) \\ \hline

        ~ & Origin & 0.2155 & 0.6578 & 0.7105 & 10  & — \\ 
        {MovieLens} & LFF-Grid & 0.2079 & 0.6747 & 0.7314 & 110  & 2.56 \\ 
         {25M}  & LFF-Random & 0.2074 & 0.6754 & 0.7322 & 231  &2.67 \\ 
        ~  & AutoAssign & 
        \underline{0.2052} & \underline{0.6792}  & \underline{0.7364} & 18  &3.25 \\ 
        ~  & AutoAssign+ & 0.2046 & \textbf{0.6817$^\star$} & \textbf{0.7386$^\star$} & 39 & 3.63 \\ \hline
        
        ~ & Origin & 0.2157 & 0.6567 & 0.7108 & 12 &— \\ 
        {MovieLens} & LFF-Grid & 0.2077 & 0.6749 & 0.7315 & 140  &2.77 \\ 
        { Latest} & LFF-Random & 0.2073 & 0.6756 & 0.7328 & 294  & 2.87 \\ 
        ~  & AutoAssign & \underline{0.2050} & \underline{0.6802} & \underline{0.7377}  & 20 & 3.56 \\ 
        ~  & AutoAssign+ & \textbf{0.2040$^\star$} & \textbf{0.6821$^\star$} & \textbf{0.7422$^\star$} & 41 & 3.87 \\ \hline
        
       ~ & Origin & 0.1914 & 0.7062 & 0.7640 & 26  &— \\ 
         {Netflix}  & LFF-Grid & 0.1819 & 0.7234 & 0.7883 & 280  &2.43 \\ 
        {Price}  & LFF-Random & 0.1818 & 0.7235 & 0.7883 & 588  & 2.45 \\ 
        ~  & AutoAssign & \underline{0.1792} & \underline{0.7293}  & \underline{0.7946} & 45  &3.27 \\
        ~  & AutoAssign+ & \textbf{0.1756$^\star$} & \textbf{0.7349$^\star$} & \textbf{0.8093$^\star$} & 87 & 4.06 \\ \hline
        \hline
    \end{tabular}}
\footnotetext{Bold denotes the highest score, and the underline indicates the best result of the baselines. * represents the significance level $p$-value $<0.05$ comparing with the best baselines.}
\end{minipage}
\end{center}
\end{table}

\begin{itemize}
\item Our method AutoAssign+ outperforms the baseline Origin in all evaluation metrics with a significant average relative improvement of 3.85\% in Accuracy. This finding supports the effectiveness of the proposed method in improving the predictive ability of the recommendation model by dynamically and automatically assigning shared embedding to low-frequency features using the critic network.

\item Baseline LFF-Grid and LFF-Random achieved a remarkable result compared to the baseline Origin (roughly 2.6\% relative improvement in average Accuracy). It demonstrates that the intuition of using shared ID embeddings to represent those low-frequency IDs is simple yet effective.

\item Our proposed method outperforms all the baselines. The LFF-Grid and LFF-Random methods require an extensive searching process and yield sub-optimal results due to several limitations. (1) They use stationary thresholds that cannot capture the complex and dynamic distribution of user and item IDs in the streaming environment. (2) These methods use only frequency as the standard of criterion and have only one shared embedding, ignoring the alteration of the distribution of ID features in space and time. In contrast, the AutoAssign+ method has a group of hierarchical candidate shared embeddings that capture the fine-grained information of features with different frequencies.
Temporally, AutoAssign+ has a fallback mechanism that captures the pattern of users who have had no interactive data for a long time but already has used a unique ID and reassigns the shared embedding to them.
\item AutoAssign+ outperforms AutoAssign on both three datasets. The Identity Agent of AutoAssign generates the candidate embeddings based solely on the input features and doesn't take into account the performance of the recommendation model. By incorporating the critic network into the optimization process, AutoAssign+ can generate more accurate and effective embeddings for the low-frequency IDs, leading to better performance of the recommendation model compared to AutoAssign.
\item The train time of grid search and random search is time-consuming compared to the Origin method, while the AutoAssign+ requires only 3.4 times of the Origin method, which is acceptable in a real industrial scenario.
\end{itemize}

\subsection{Cold-Start Stage}
\begin{figure}
\centering
\includegraphics[width=0.9\linewidth]{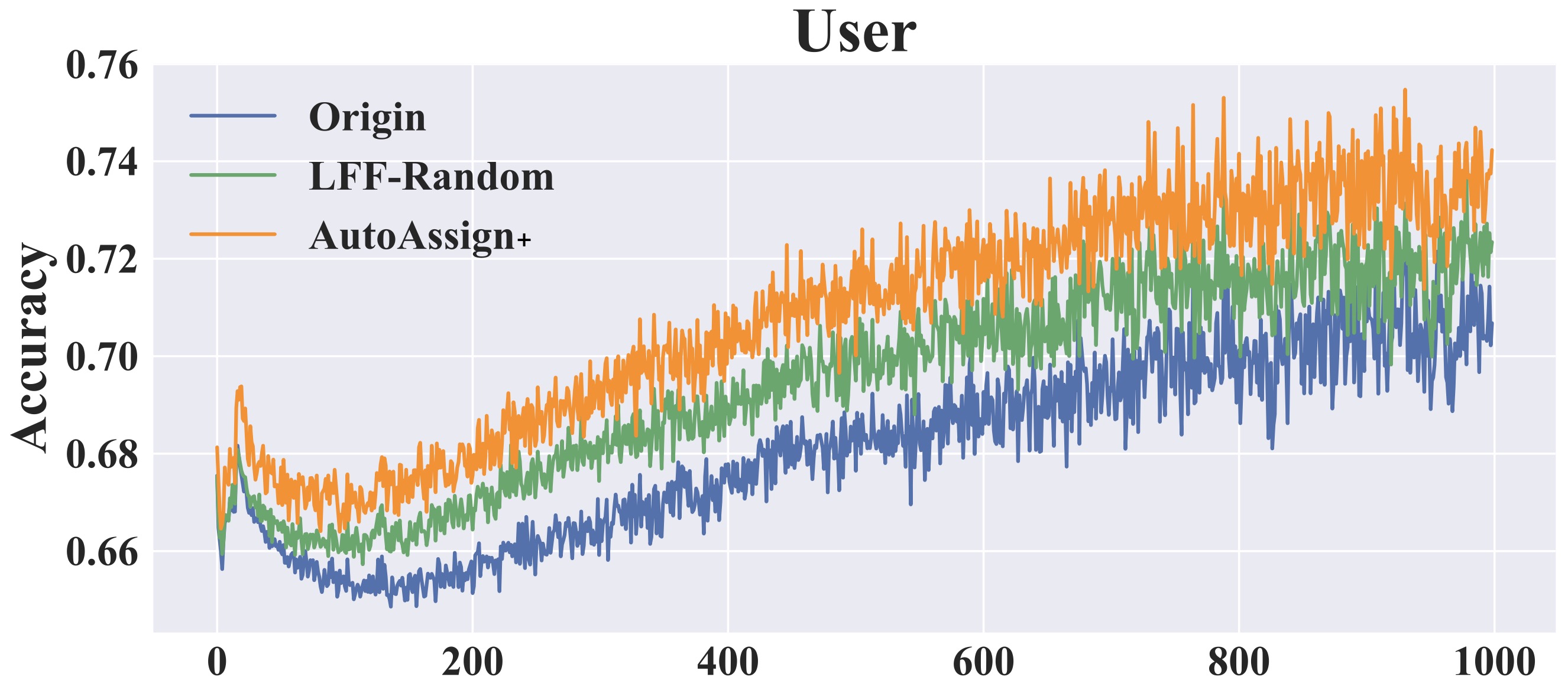}
\includegraphics[width=0.9\linewidth]{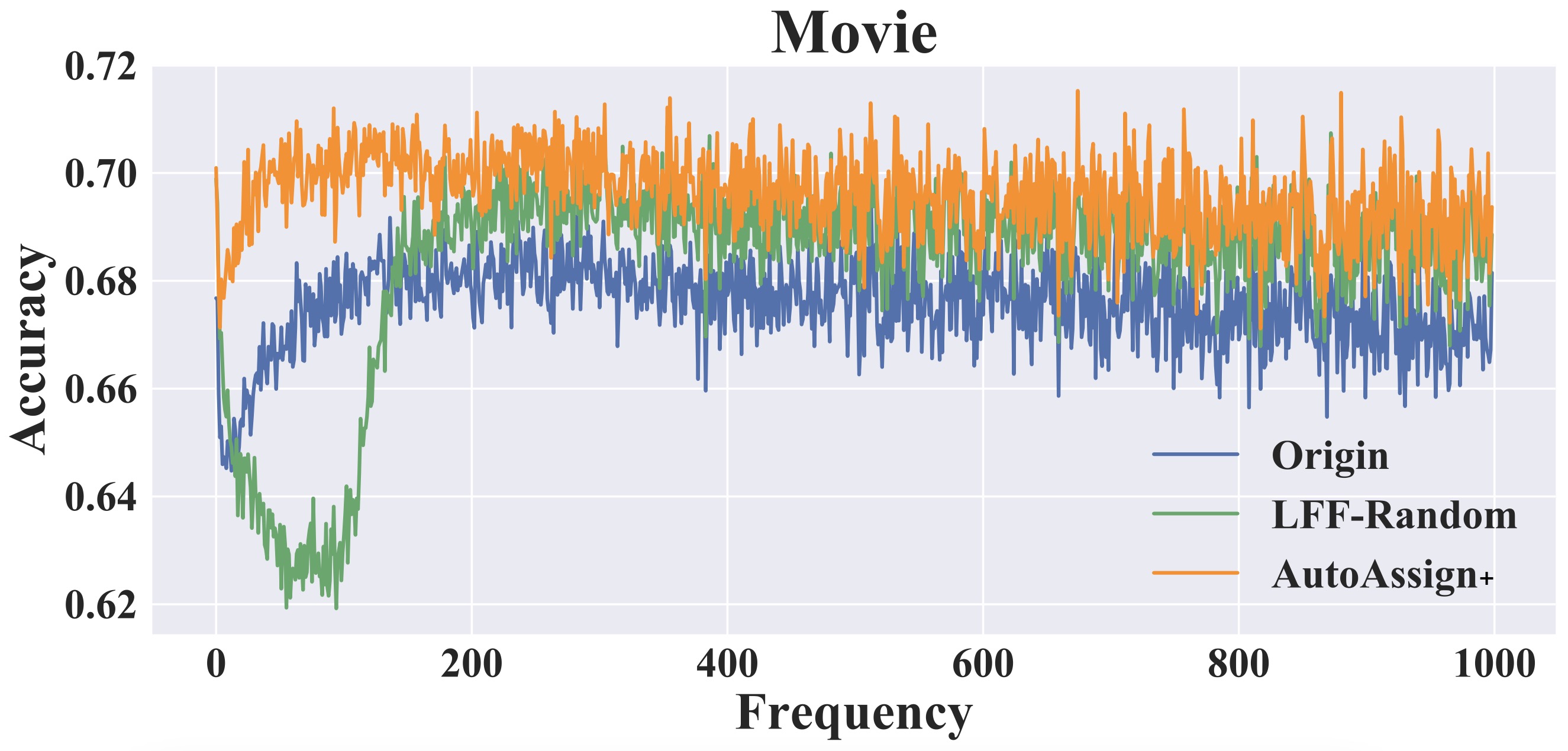}
\vspace{-1mm}
\caption{Accuracy over different frequencies.}
\label{cold}
\vspace{-5mm}
\end{figure}

This experiment compares the performance of our proposed AutoAssign+ method with other baselines in terms of the cold-start stage in recommendation systems. As shown in Fig. \ref{cold}, we choose the MovieLens Latest dataset, and the average accuracy of different user and movie frequencies was plotted, varying from 0 to 1,000. The results showed that our proposed AutoAssign+ method consistently outperformed other baselines in terms of accuracy. Specifically, in the user part, the gap between our method and other baselines was more significant, with an average accuracy improvement of 1.1\% compared to LFF. In the movie part, the baseline LFF suffered a dramatic downward trend in the beginning stage due to the low number of movies, while AutoAssign+ reached a stable and much higher performance. The reason for this is that a fixed shared embedding threshold makes all movies use one ID embedding, causing a severe deviation. However, AutoAssign+ can dynamically make different judgments for each movie by minimizing the different losses that result from using the shared ID and smoothly avoiding the deviation caused by the single frequency information, even when using frequency information similarly.

\subsection{Ablation Study}
An ablation experiment was conducted on MovieLens Latest dataset to demonstrate how each component of AutoAssign+ contributed to the overall performance. The method "No Descend" involved reducing the output of the agent network to only two actions: "Ascend" and "Unchanged," while "Single SE" meant setting the number of shared embeddings of both userID and itemID to only 1. The results in Table \ref{ablation} showed that both AutoAssign (No Descend) and AutoAssign (Single SE) produced a minor gap from AutoAssign but still achieved a superior performance to the baseline LFF. However, AutoAssign (Single SE + No Descend) performed worse than LFF. The reward of the RL-based Identity agent was based on the last five prediction losses of each ID, and the behavior of different IDs dynamically changed in both spatial and temporal manners. In such a case, the use of both uni-directional action and a single shared embedding would make the agent's decision excessively judgmental, generating the same deviation as LFF. Therefore, the hierarchical group of shared embedding and descend action had to be used simultaneously to achieve a better result. The critic network in AutoAssign+ further improved the model performance by optimizing the Identity Agent's policy, enhancing the stability and efficiency of the framework.

\begingroup
\setlength{\tabcolsep}{4pt} 
\renewcommand{\arraystretch}{1.4}
\begin{table}[!ht]
    \centering
    \vspace{-3mm}
     \caption{Effectiveness of Different Components}
     \label{ablation}
     \vspace{-1mm}
    \begin{tabular}{ c c c c }
    \hline
        Method & MSE loss & Accuracy & Improve \\ \hline
        Origin & 0.2155 & 0.6578 & — \\ 
        AutoAssign(No Descend) & 0.2059 & 0.6785 & 3.15\% \\ 
        AutoAssign(Single SE) & 0.2056 & 0.6793 & 3.26\% \\ 
        AutoAssign(Single SE + No Descend) & 0.2082 & 0.6732 & 2.34\% \\ 
        AutoAssign & 0.2052 & 0.6801 & 3.40\% \\ 
        AutoAssign+ & 0.2047 & 0.6819 & 3.66\% \\
        \hline
    \end{tabular}
  \vspace{-4.9mm}
\end{table}
\endgroup

\subsection{Practicality Comparison}
\begin{figure}[h]
\vspace{-3mm}
  \centering
  \includegraphics[width=\linewidth]{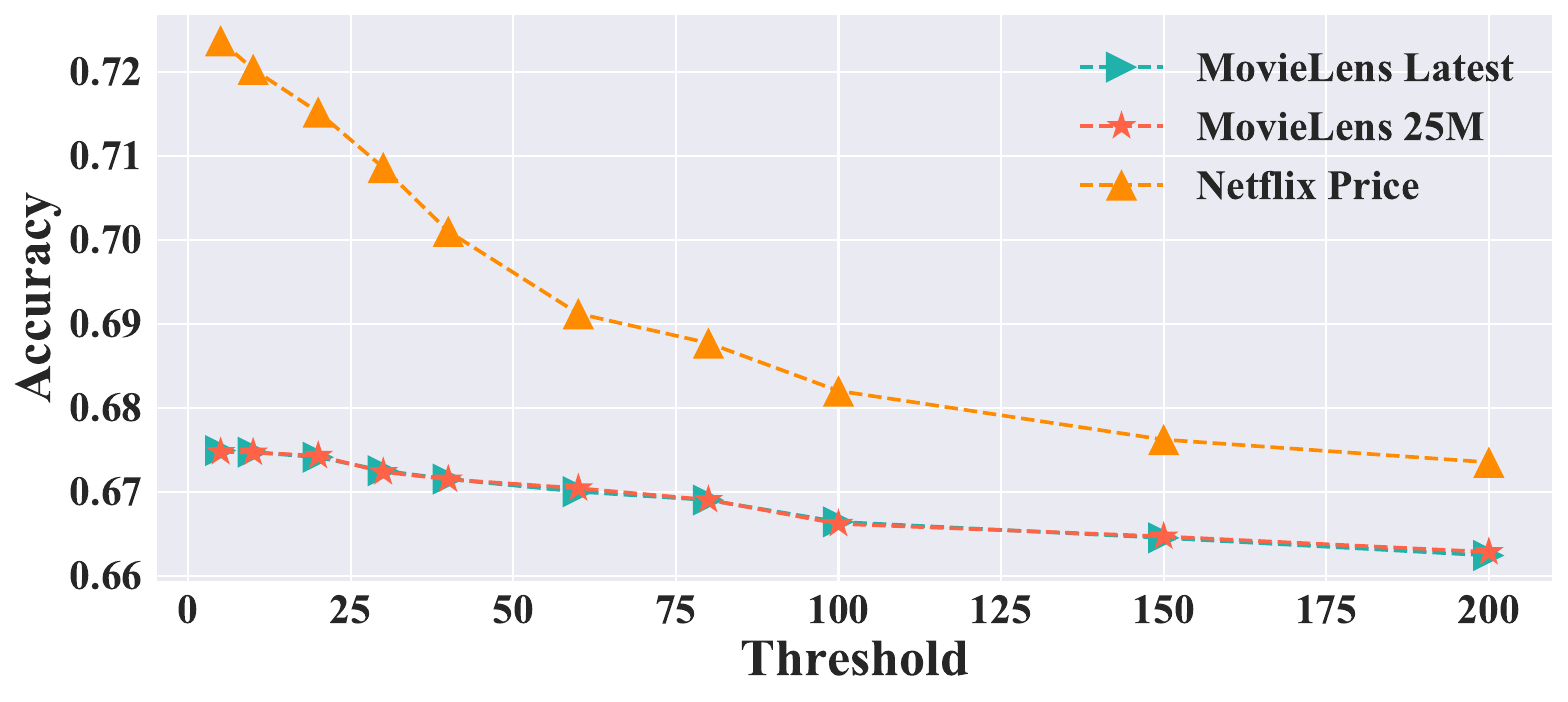}
  \vspace{-7mm}
  \caption{Grid Search of LFF Threshold.}
  \label{grid}
  \vspace{-5mm}
\end{figure}

\begin{figure}
\centering
\includegraphics[width=\linewidth]{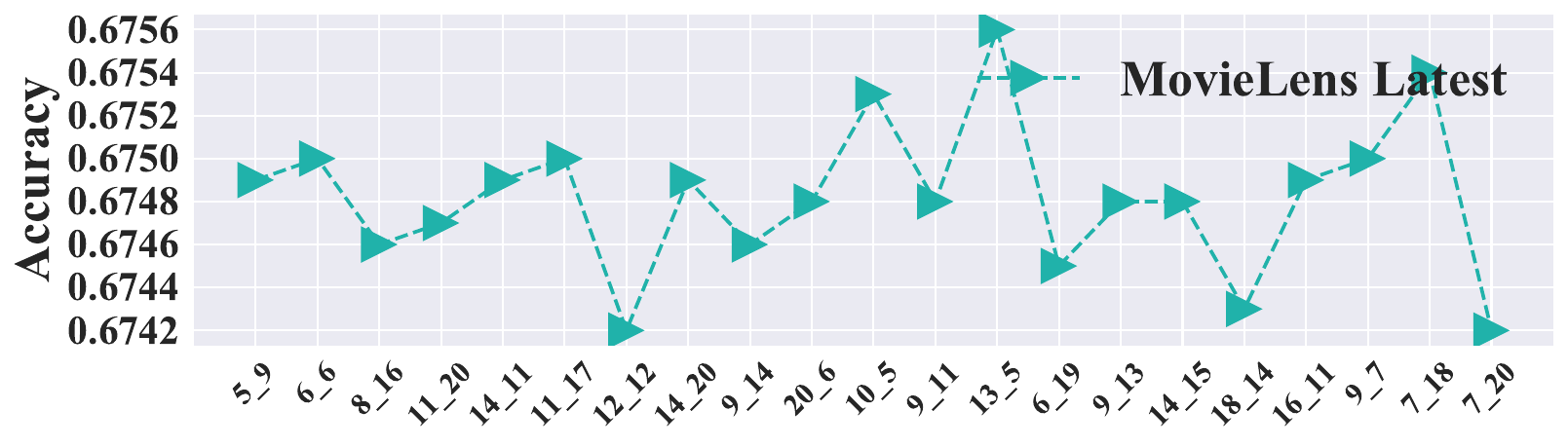}
\includegraphics[width=\linewidth]{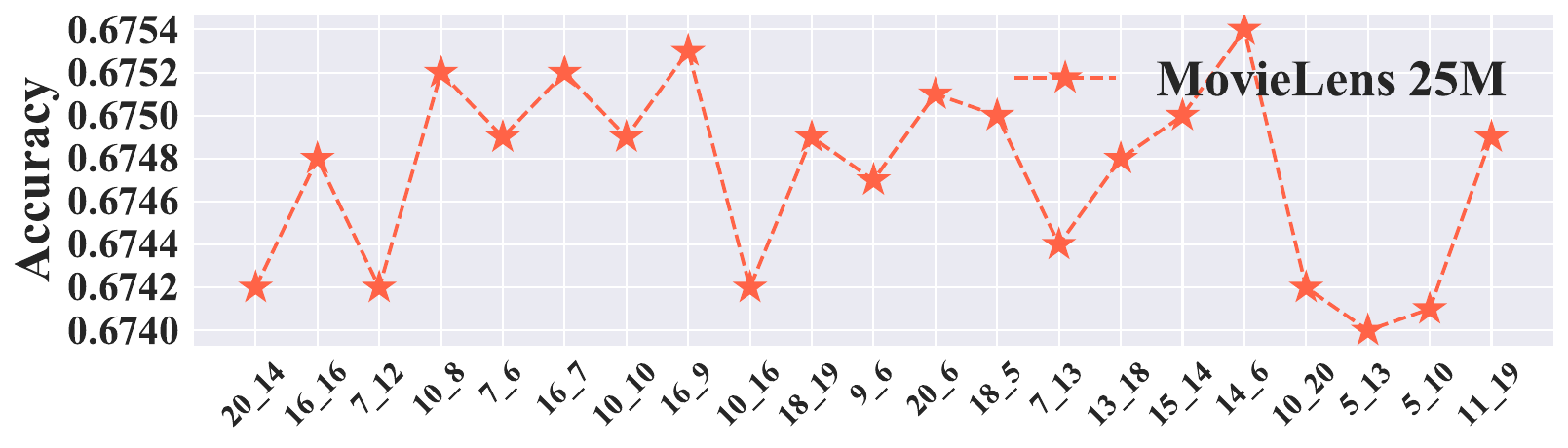}
\includegraphics[width=\linewidth]{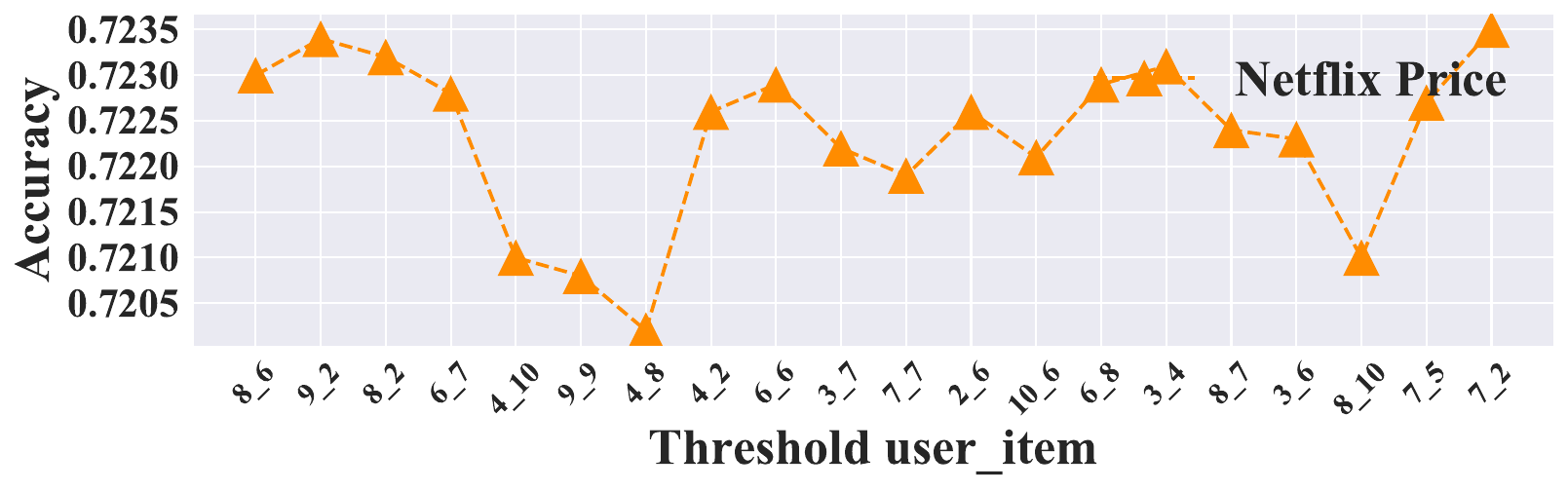}
\vspace{-5mm}
\caption{Random Search of LFF Threshold.} 
\label{random}
\vspace{-5mm}
\end{figure}

To ensure fairness in the evaluation of the low-frequency filter, we apply grid search and random search methods to optimize its performance. As shown in Fig. \ref{grid}, the frequency thresholds for the user and movie were selected from a range of 5 to 200 with an interval of 10 in the grid search, and the same threshold was applied to all three datasets. We observed that the performance of the low-frequency filter decreased monotonically when the frequency threshold was set to 20 or higher. To further improve the effectiveness of the low-frequency filter, we conducted a random search for different threshold combinations on the user and movie sides in the interval of 5-20. The threshold group with the highest Accuracy was selected as the experimental group for overall performance, as presented in Fig. \ref{random}. Also, note that in Table  \ref{overall}, the train time of AutoAssign is much faster than grid search and random search, proving our model's efficiency in real-world recommendations.

\subsection{Parameter Reduction}
Reducing memory consumption is a crucial factor in designing efficient recommender systems, especially for online platforms. The embedding layer is one of the most significant contributors to the total number of parameters in deep recommender models. The proposed Automated Shared Embedding approach in AutoAssign+ aims to reduce memory consumption by assigning shared embeddings to low-frequency IDs, thus avoiding unnecessary allocation of unique embeddings to these IDs.
Table \ref{para} presents the parameter usage of the embedding layer in AutoAssign+ and the deduction ratio achieved by using the shared embedding approach. The \textit{Deduction} column shows the percentage of parameters saved in different datasets, and the \textit{Total} column shows the overall deduction ratio. The results demonstrate that the shared embedding approach reduces the total number of parameters in the embedding layer by 20\% to 30\%, depending on the dataset. At the same time, the model's accuracy is improved due to the enhanced representative ability of the shared embeddings. Therefore, the Automated Shared Embedding approach in AutoAssign+ can reduce memory consumption without compromising the model's accuracy.

\vspace{-4mm}
\begingroup
\setlength{\tabcolsep}{4pt} 
\renewcommand{\arraystretch}{1.4}
\begin{table}[h]
    \centering
    \caption{Parameter Deduction on AutoAssign+}
    \vspace{-1mm}
    \label{para}
    \begin{tabular}{ c c c c c }
    \hline
        Dataset(field) & Origin Para & AutoAssign+ Para & Deduction & Total \\ \hline
        ML-Latest(user) & 35,840,000 & 25,553,937 & 28.7\% & 31.3\% \\ 
        ML-Latest(movie) & 7,424,000 & 4,164,872 & 43.9\%  & ~ \\ \hline
        ML-25M(user) & 20,736,000 & 16,153,349 & 22.1\% & 23.1\% \\ 
        ML-25M(movie) & 7,936,000 & 5,833,016 & 26.5\% & ~ \\ \hline
    \end{tabular}
    \vspace{-6mm}
\end{table}
\endgroup

\begin{figure}[ht]
\centering
\vspace{-2mm}
\includegraphics[width=0.98\linewidth]{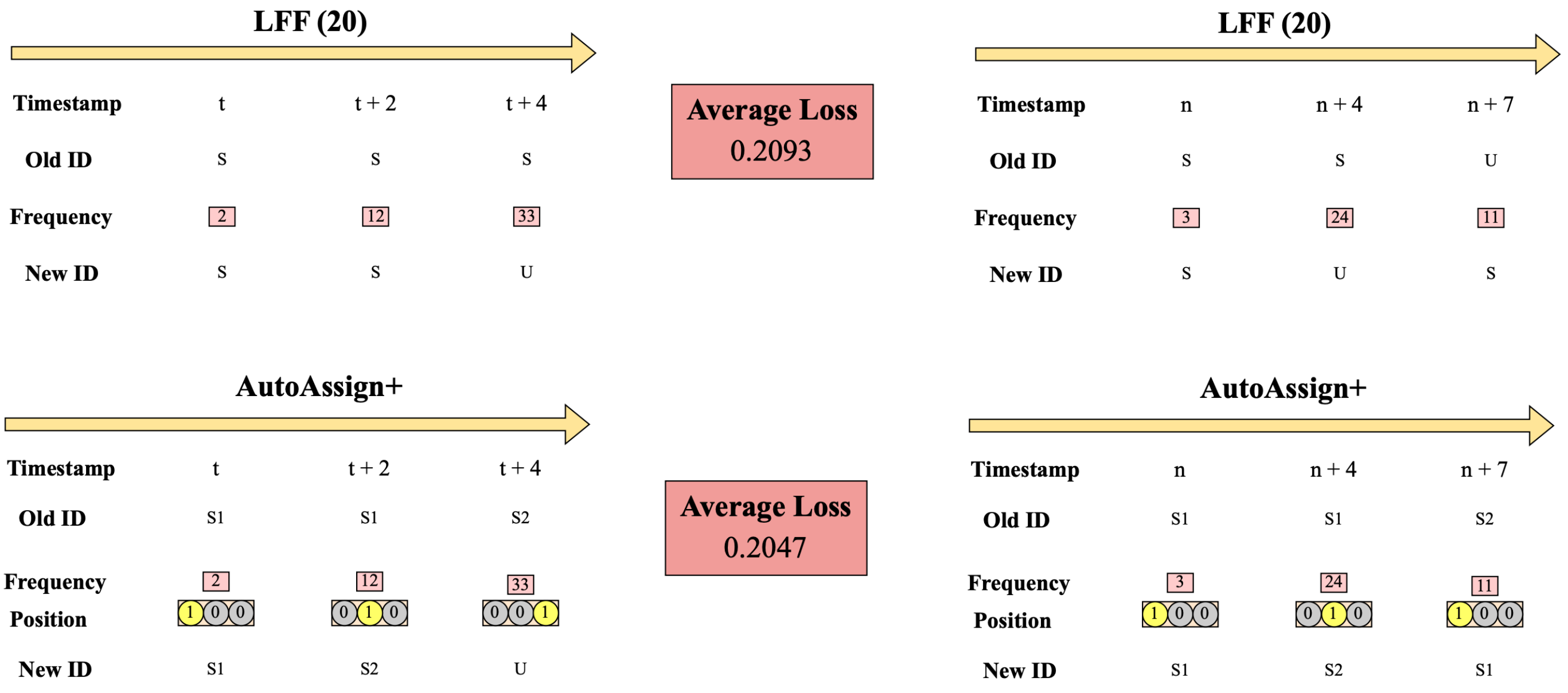}
\vspace{-1mm}
\caption{Case Study on MovieLens-25M.}
\label{cs}
\vspace{-7mm}
\end{figure}

\subsection{Case Study}
In this subsection, we present a comparison between our proposed AutoAssign+ method and the LFF baseline to illustrate the effectiveness of our approach. On the left-hand side instance, we consider a new user from the MovieLens-25M dataset, whose frequency increases to 12 at timestamp $t+2$, which is below the threshold of 20. In this case, LFF assigns a shared ID to the user. However, at timestamp $t+4$, the user frequency increases to 33, which is above the threshold, leading LFF to assign a specific user ID to this user. In contrast, AutoAssign+ adopts a different ID assignment strategy. At timestamp $t+2$, when the user frequency increases to 12 with a position shift from $p_1$ to $p_2$, AutoAssign+ assigns a shared ID with a higher hierarchy to the user. This hierarchical candidate shared ID setting with reinforcement learning powered embedding assignment results in a reduction in the average loss from 0.2093 to 0.2047, which highlights the efficiency of AutoAssign+.

\section{Related Work}
\subsection{Cold-Start problem}
The purpose of the recommender system is to recommend a set of items that the user may be interested in. However, if the user interaction data is limited, the performance of the recommendation model will be significantly reduced, which is called the cold-start problem. Cold-start problem is a ubiquitous and challenging problem in personalized recommendation, where extensive studies have been done concerning this issue \cite{FENG2021106732,wahab2022federated,jeevamol2021ontology,LIU2023110579,wang2020cdlfm}. Some content-based methods \cite{cold1} make relevance between zero shot-learning and cold-start recommendation and propose a low-rank linear auto-encoder to solve the cold-start problem using the user's auxiliary information. Internal and contextual attention networks (ICAN) \cite{cold3} strengthen the interaction of the feature domain and use auxiliary information among multiple queues to get a better cold-start performance in the matching stage. Meta-learning is a common method to learn the global and general information for pertinent tasks and serves to speed up the initialization of new relative tasks. MetaEmb \cite{meta1} and MeLU \cite{melu} apply Model-Agnostic Meta-Learning, where MetaEmb trains a generator to initialize embedding, and MeLU learns the initialization parameters of the whole model to solve the cold-start problem. MWUF \cite{meta4} uses the average pooling of all items to initialize the new item embedding and uses two meta-network to enhance their representation. These methods mentioned above have their own limitations and are discussed in the introduction section.

\subsection{Reinforcement Learning Based Recommendation}
Numerous studies have explored the integration of Reinforcement Learning (RL) with Recommender Systems (RS)~\cite{liu2023multi,afsar2021reinforcement,wang2022surrogate,zhang2022multi, zhao2017deep, zhao2018deep, zhao2021dear,liu2022redrl}. Instead of optimizing immediate user feedback similar to traditional learning-to-rank approaches ~\cite{liu2009learning}, RL-based Recommender Systems seeks to optimize the cumulative reward function, which estimates multiple rounds of interactions between the recommendation policy and the user response environment. Specifically, the problem of sequential user-system interactions can be formulated as a Markov Decision Process (MDP)~\cite{shani2005mdp}. 
Various RL solutions have been investigated under this formulation, including tabular-based methods that store and update a table that represents the estimated value or quality of each state-action pair ~\cite{joachims1997webwatcher,mahmood2007learning,moling2012optimal}, value-based methods utilized to assess the effectiveness of a specific action or state~\cite{taghipour2007usage,zheng2018drn,zhao2018recommendations,ie2019slateq}, policy gradient methods that optimize the recommendation policy based on long-term reward~\cite{chen2019top,chen2019large}, and actor-critic methods ~\cite{liu2018deep} that simultaneously learn an action evaluator network and action generator, which is based on policy gradient ~\cite{peters2008natural,bhatnagar2007incremental,degris2012model,10.1145/3539618.3591717}. 
Some of the major difficulties in employing RL for recommender systems include the vast state and action space~\cite{dulac2015deep,liu2020state}, uncertainty in the user environment~\cite{ie2019recsim,zhao2019deep}, exploration effectiveness and efficiency ~\cite{chen2021user}, and creating suitable reward function that caters to diverse behaviors~\cite{zou2019reinforcement}. Our work addresses reward function design while simultaneously enhancing the agent's performance by utilizing the actor-critic framework.

\section{Conclusion}
In this study, we analyze the cold-start problem in streaming deep recommender systems, which results in poorly trained ID embeddings and reduces prediction performance, as well as leads to unnecessary memory usage in the model. To address these issues, we propose a framework called Automatic Shared Embedding Assignment Plus (AutoAssign+), which includes a critic network-enhanced Identity Agent that can automatically and field-wisely assign shared IDs to low-frequency IDs. AutoAssign+ reduces the human effort required in the time-consuming search process and expert knowledge and takes action to improve performance and reduce the model's parameter usage by 20\%-30\%. The proposed framework is independent of the inference layer, making it easily applicable to various existing recommendation models that use an embedding lookup layer. We demonstrate the effectiveness of AutoAssign+ in addressing the cold-start problem and the practicality of their approach through extensive experiments on three popular datasets. However, the selection of actions (ascend/unchanged/descend) made by the actor network is somewhat opaque and not easily interpretable, which makes it difficult to gain a clear understanding of how the system is making decisions. This is one of the directions for further improvement.

\section*{ACKNOWLEDGEMENT}
This research was partially supported by Huawei (Huawei Innovation Research Program), APRC - CityU New Research Initiatives (No.9610565, Start-up Grant for New Faculty of City University of Hong Kong), CityU - HKIDS Early Career Research Grant (No.9360163), Hong Kong ITC Innovation and Technology Fund Midstream Research Programme for Universities Project (No.ITS/034/22MS), SIRG - CityU Strategic Interdisciplinary Research Grant (No.7020046, No.7020074). We thank MindSpore \cite{mindspore} for the partial support of this work, which is a new deep learning computing framework.

\bibliography{ref}

\end{document}